\documentclass[apj]{emulateapj}
\usepackage{lscape}
\usepackage{apjfonts}
\usepackage{graphicx}

\shorttitle{Photometric Metallicity of the Virgo Overdensity}
\shortauthors{An et~al.}

\begin{document}
\title{A Photometric Metallicity Estimate of the Virgo Stellar Overdensity}

\author{Deokkeun An\altaffilmark{1},
Jennifer A.\ Johnson\altaffilmark{2},
Timothy C.\ Beers\altaffilmark{3},
Marc H.\ Pinsonneault\altaffilmark{2},\\
Donald M.\ Terndrup\altaffilmark{2,4},
Franck Delahaye\altaffilmark{5,6,7},
Young Sun Lee\altaffilmark{3},\\
Thomas Masseron\altaffilmark{2},
Brian Yanny\altaffilmark{8}
}
\altaffiltext{1}{Infrared Processing and Analysis Center, California Institute
of Technology, Mail Stop 100-22, Pasadena, CA 91125; deokkeun@ipac.caltech.edu.}
\altaffiltext{2}{Department of Astronomy, Ohio State University, 140 West 18th
Avenue, Columbus, OH 43210.}
\altaffiltext{3}{Department of Physics \& Astronomy and JINA: Joint Institute
for Nuclear Astrophysics, Michigan State University, E. Lansing, MI 48824.}
\altaffiltext{4}{Division of Astronomical Sciences, National Science Foundation,
4201 Wilson Blvd., Arlington, VA 22230.}
\altaffiltext{5}{Service d'Astrophysique, CEA/DSM/IRFU/SAp, CEA Saclay, 91191
Gif-sur-Yvette Cedex, France.}
\altaffiltext{6}{Centre Lasers Intenses et Applications (CELIA), 351 Cours de
la Lib\'eration, 33405 Talence Cedex, France.}
\altaffiltext{7}{LERMA, Observatoire de Paris, CNRS, Universit\'e Paris Diderot,
5 place Jules Janssen, 92190 Meudon, France.}
\altaffiltext{8}{Fermi National Accelerator Laboratory, P.O. Box 500, Batavia,
IL 60510.}

\begin{abstract}
We determine photometric metal abundance estimates for individual main-sequence
stars in the Virgo Overdensity (VOD), which covers almost $1000\ \deg^2$ on the
sky, based on a calibration of the metallicity sensitivity of stellar isochrones
in the $gri$ filter passbands using field stars with well-determined spectroscopic
metal abundances. Despite the low precision of the method for individual stars,
we derive ${\rm [Fe/H]} = -2.0\pm0.1\ {\rm (internal)} \pm0.5\ {\rm (systematic)}$
for the metal abundance of the VOD from photometric measurements of $0.7$ million
stars in the Northern Galactic hemisphere with heliocentric distances from $\sim10$~kpc
to $\sim20$~kpc. The metallicity of the VOD is indistinguishable, within
$\Delta {\rm [Fe/H]} \le 0.2$, from that of field halo stars covering the same distance
range. This initial application suggests that the SDSS $gri$ passbands can be used
to probe the properties of main-sequence stars beyond $\sim10$~kpc, complementing
studies of nearby stars from more metallicity-sensitive color indices that involve
the $u$ passband.
\end{abstract}


\section{Introduction}

The structure, chemistry, and kinematics of the stellar halo of the Galaxy, with its
predominantly old and metal-poor populations, collectively preserve a detailed record
of our Galaxy's formation in the early universe \citep[e.g.,][]{eggen:62,searle:78}.
Thanks to large-area surveys such as the Sloan Digital Sky Survey
\citep[SDSS;][]{york:00,edr,dr1,dr2,dr3,dr4,dr5,dr6,dr7}, recent studies have revealed
that the halo is marked by numerous stellar substructures. The presence of these lumpy
and complex substructures (both in real and phase space) are in qualitative
agreement with models for the formation of the stellar halo through the hierarchical
merging and accretion of low-mass subhalos \citep[e.g.,][]{bullock:05}.

Among the various substructures discovered to date, the Virgo Overdensity (VOD) is one
of the most striking.  It was discovered as a stellar overdensity of main-sequence
stars in SDSS; starcounts in the region toward Virgo are enhanced by a factor of two
above the background stellar distribution \citep{juric:08}. The overdensity covers
almost $1000\ \deg^2$, and it appears to span a wide range of heliocentric distances
of $\sim5$~kpc--$20$~kpc. The overdensity seems to be associated with clumps of RR Lyrae
stars \citep{vivas:01,duffau:06} and turn-off stars \citep{newberg:02}, but it is less
likely to be connected with the leading tidal tail of the Sagittarius dwarf galaxy
\citep{newberg:07}.

At present, metallicity estimation from broadband photometry is the only practical
means of obtaining metal abundances for a large number of faint objects such as those
in the VOD. Such methods are based on the relative sensitivity of stellar colors to
photospheric abundances over a wide wavelength baseline. The clear advantage of using
a photometric metallicity technique is the efficiency of estimating metallicities for
individual main-sequence stars, which are the most plentiful and representative sample
of stellar populations.

\citet{ivezic:08a} constructed photometric metallicity relations in the $u - g$
vs.\ $g - r$ plane using SDSS filter passbands, and studied the abundance structures
of the Galaxy with an accuracy of $\sim0.2$~dex at $g < 17$. This approach is similar
to the traditional $UBV$ method \citep[e.g.,][]{carney:79}, which relies on the strong
dependence of $U$-band
magnitudes on metal abundance. However, the $u$-band photometry in SDSS is limited
to $u \approx 22$ (99\% detection limit). This, and the greatly deteriorating errors
in SDSS $u$-band magnitudes near the faint limit, restricts photometric metallicity
estimates to stars at $r \la 20.8$, an insufficient depth to fully explore the VOD
\citep[see Fig.~37 in][]{juric:08}.

In this letter we overcome the limitations of the $u$-band photometry in SDSS by
exploring less metallicity-sensitive, but better-determined, color indices in the $gri$
passbands.\footnote{The $z$-band data were not used due to the bright
survey limit in SDSS for this filter.} Turn-off stars in globular clusters have
$M_r \sim 4$, so the SDSS survey limit in $gri$ ($95\%$ completeness limit at $r = 22.1$)
allows us to probe the halo out to $\sim20$~kpc using stars that are $\sim2$~mag below
the main-sequence turn-off ($\sim0.6\ M_\odot$).

\section{Method}

\subsection{Photometry}

We employed SDSS photometry from DR7 \citep{dr7}. SDSS measures the
brightnesses of stars with accurate astrometric positions \citep{pier:03},
using a dedicated 2.5-m telescope \citep{gunn:06} in five broadband filters
\citep[$ugriz$;][]{edr}, on 6 columns of CCDs
\citep{gunn:98}, under photometric conditions \citep{hogg:01}. Photometric
calibration is carried out using observations of stars in the secondary patch
transfer fields \citep{tucker:06}, based on the \citet{smith:02} sample of
standard stars.

The rms photometric precision is $0.02$~mag for sources not limited by photon
statistics, and the photometric calibration is accurate to $\sim2\%$ in the $gri$
bands, and $\sim3\%$ in $u$ and $z$ \citep{ivezic:03,ivezic:04,an:08}.
We used photometry of stellar objects \citep[identified as {\tt STAR} in the
standard SDSS photometric pipelines;][]{lupton:02} that were detected (at the
$5\sigma$ level) in all of the $gri$ passbands. The observed magnitudes are
corrected for extinction adopting reddening values in the \citet{schlegel:98}
dust maps and the extinction coefficients given by \citet{an:09}.

\subsection{Photometric Metallicity}

The isochrones in \citet{an:09} were used to determine photometric
metallicity (${\rm [Fe/H]_{phot}}$) estimates based on color-color relations for
main-sequence stars. In the following analysis, we adopted the same $\alpha$-element
enhancement scheme as in \citet{an:09}, motivated by the observed behavior of
these elements among field dwarfs and cluster stars from spectroscopic studies
\citep[e.g.,][]{venn:04,kirby:08a}:
[$\alpha$/Fe]$ = +0.4$ at [Fe/H]$ = -3.0$,
[$\alpha$/Fe]$ = +0.3$ at [Fe/H]$ = -2.0, -1.5, -1.0$,
[$\alpha$/Fe]$ = +0.2$ at [Fe/H]$ = -0.5$, and
[$\alpha$/Fe]$ = +0.0$ at [Fe/H]$ = -0.3, -0.2, -0.1, +0.0, +0.1, +0.2, +0.4$.
A linear interpolation was used in this metallicity grid to obtain isochrones
at intermediate [Fe/H] values. We adopted an age of 12.6~Gyr for ${\rm [Fe/H]}
\leq -1$, and 4.6~Gyr at ${\rm [Fe/H]} \geq 0$, and a linear interpolation
between these two values. To expedite the computational process, we derived
$5^{th}$-order polynomials to describe the color-magnitude
relations, with intervals in abundance of ${\rm \Delta [Fe/H]} = 0.01$~dex.

To reliably estimate a photometric metal abundance, it is
necessary to use stellar models that match cluster main-sequences and give
internally consistent distances from multiple color-magnitude
diagrams (CMDs). However, despite improvements in theoretical models, there
still exist small but significant mismatches between calculated isochrone colors
and the best photometry in well-studied open clusters \citep{pinsono:04,an:07a,an:07b}.
Therefore, we applied empirical corrections on theoretical color-$T_{\rm eff}$
relations for ${\rm [Fe/H]} \geq -0.8$ to match the observed main sequence of
the solar-metallicity cluster M67 \citep{an:09}. Constant correction factors
were adopted at ${\rm [Fe/H]} \geq 0$, and a linear ramp was used between
${\rm [Fe/H]} = -0.8$ and ${\rm [Fe/H]} = 0$, so that the correction becomes
zero at ${\rm [Fe/H]}=-0.8$ and below, as the models are in good agreement with
the data for globular clusters \citep{an:09}.

\begin{figure}
\epsscale{1.2}
\plotone{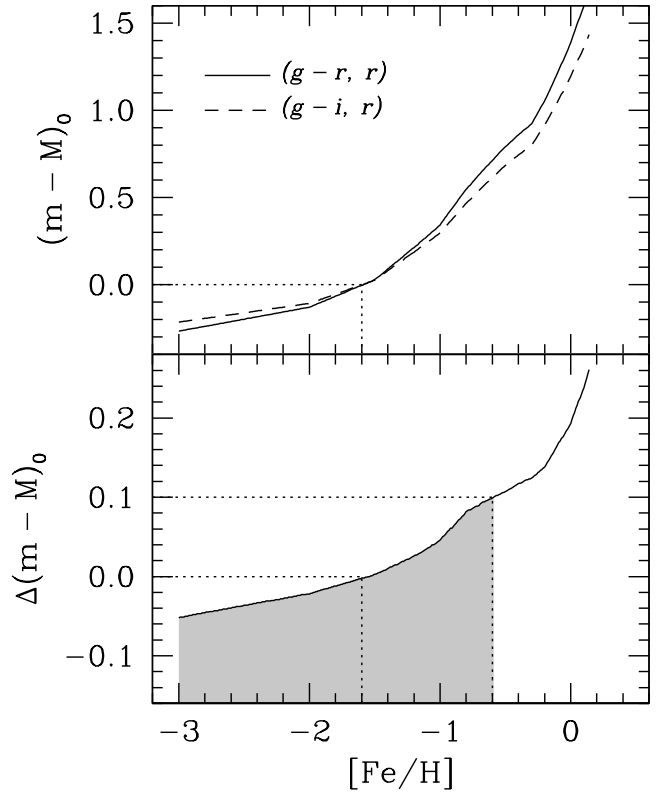}
\caption{Illustration of the photometric metallicity estimation technique.
{\it Top}: Distance moduli from two CMDs as a function of a metallicity. For
this example, ${\rm [Fe/H]} =-1.6$ was assumed for the true metallicity of
a star. {\it Bottom}: Difference in distance modulus from the two CMDs. A
photometric metallicity is defined as the [Fe/H] that results in the same
distance modulus from the two CMDs. The gray region shows the error bound
for a $\sim2\%$ photometric color error. Note that only the upper [Fe/H] bound
is defined in this example.
\label{fig:fehdist}}
\end{figure}

A photometric metallicity \citep{an:07b} was computed for each star by requiring
distances from main-sequence fitting to be the same from two different CMDs,
with $g - r$ and $g - i$ as color indices and $r$ as a luminosity index
(Fig.~\ref{fig:fehdist}). We searched the entire [Fe/H] grid from $-3.0$ to
$+0.4$ to find [Fe/H] with a minimum $\chi^2$ value, defined as
\begin{equation}
\chi^2 = \frac{(\mu_{g-r} - \bar{\mu})^2}{\sigma_{\mu_{g-r}}^2}
     + \frac{(\mu_{g-i} - \bar{\mu})^2}{\sigma_{\mu_{g-i}}^2},
\end{equation}
for each star. Here, $\mu$ and $\sigma_{\mu}$ are the distance modulus and its
error for each CMD, respectively. The quantity $\bar{\mu}$ is a weighted average
distance modulus from the $(g - r, r)$ and $(g - i, r)$ CMDs. Since only three passbands
are considered here, the problem is reduced to the traditional manner of
determining metal abundances from a color-color diagram. In some cases, a star
becomes bluer than the main-sequence turnoff as the metallicity increases,
either due to a large photometric error or a younger age than our assumed values
in the models. If a minimum $\chi^2$ was not found, ${\rm [Fe/H]_{phot}}$ was
estimated based on a mean relation between [Fe/H] and $\Delta (m - M)_0$.

\subsection{Accuracy of Photometric Metallicity Determinations}

Photometric metallicity estimates become insensitive for very metal-poor stars,
as illustrated in Figure~\ref{fig:fehdist}. Therefore, ${\rm [Fe/H]_{phot}}$ could
be biased due to systematic effects such as
photometric errors, unresolved binary stars, and dust extinction, which alter
the differential color estimates at a small level.  We tested our photometric
metallicity estimates using field stars with well-measured spectroscopic estimates,
as discussed below, and adjusted ${\rm [Fe/H]_{phot}}$ to correct for the bias.

\begin{figure}
\epsscale{1.2}
\plotone{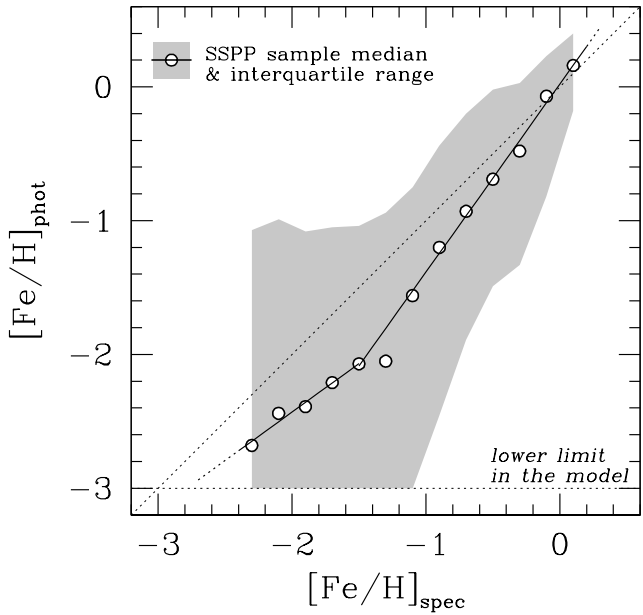}
\caption{Comparison between spectroscopic and photometric estimates of [Fe/H].
Open circles are median ${\rm [Fe/H]_{phot}}$ for the SSPP field star sample with
available spectroscopic metallicities; the gray region represents interquartile
ranges. Solid lines are piecewise linear fits to the open circles.
\label{fig:cluster}}
\end{figure}

Figure~\ref{fig:cluster} shows the comparison with low-resolution ($R\sim 2000$)
spectroscopic measurements for field dwarfs \citep{carollo:07,yanny:09} from
the most recent version (Data Release planned in December 2010)
of SSPP \citep[SEGUE Stellar Parameter Pipeline;][]{lee:08a,lee:08b}.
This version of the SSPP partially solves the under- and over-estimation of metallicity
at higher and lower ends of [Fe/H] that the earlier version (DR7) showed.
Comparisons are shown for 46,983 stars, after applying cuts at ${\rm S/N} > 20$ and
$\log{g} \geq 4.2$ to select dwarfs with good spectroscopic abundance measurements.
We further restricted the sample to those with $| 1.4(g - r)_0 - (g - i)_0 | \leq 0.1$.
Open points show a median metallicity in ${\rm [Fe/H]}=0.2$~dex bins; the gray
region represents interquartile ranges. Because the metallicity sensitivity
essentially disappears below the lower limit of our metallicity grid, a significant
number of stars are found at ${\rm [Fe/H]_{phot}} = -3.0$.

The large uncertainty in ${\rm [Fe/H]_{phot}}$ is mainly due to photometric errors.
We performed artificial star tests by generating stars from
the isochrones with Gaussian errors of $0.02$~mag in $r$, $g - r$, and $g - i$. The
interquartile range from the test showed a reasonable agreement with those for the
SSPP sample, indicating that the observed dispersion is at least in part due to the
$\sim2\%$ photometric error in SDSS. Despite the large uncertainty in
${\rm [Fe/H]_{phot}}$ for individual stars, a meaningful constraint on a median
${\rm [Fe/H]_{phot}}$ can be made by applying the technique to a large number of
stars ($N\sim500$, see below).

For the SSPP sample in Figure~\ref{fig:cluster}, we estimated the error in the median
of $\la0.1$~dex at ${\rm [Fe/H]_{spec}} > -2$ in each [Fe/H] bin; this was done by
computing the median absolute deviation (MAD)\footnote{${\rm MAD} \equiv 1.483\ {\rm median}
(|x_i - {\rm median}(x_i)|)$.} from stars with ${\rm [Fe/H]_{phot}}$ above the median,
divided by $\sqrt{N}$ (although ${\rm [Fe/H]_{phot}}$ does not strictly follow a normal
distribution). We explored the effects of differing stellar ages ($\sigma = 20\%$),
[$\alpha$/Fe] ratios ($\sigma \approx 0.1$~dex at ${\rm [Fe/H]} \la -1$), and dust
extinctions ($\sigma = 20\%$), but they change
${\rm [Fe/H]_{phot}}$ by $\la0.1$~dex.

As shown in Figure~\ref{fig:cluster}, our photometric solution underestimates [Fe/H] by
as much as $\Delta {\rm [Fe/H]} \sim 0.5$~dex for ${\rm [Fe/H]} \la -1$ with respect to
the SSPP results. One reason for this offset is likely due to the presence of
unresolved binary populations in the sample, which will have a portion of their light
from a cooler secondary \citep{an:07b}. To evaluate the effect of binaries, we performed
artificial star tests with binaries generated from a flat mass function for the secondaries,
and found that the $\Delta {\rm [Fe/H]} \sim 0.5$~dex offset can be explained with
a $\sim40\%$ binary fraction. On the other hand, the difference could also be induced
by a systematic color offset in the isochrones. Although we found a good agreement of
the models with globular cluster data within the errors \citep{an:09}, a small
color offset ($\la0.01$~mag) in the models can still change the photometric metallicity
estimate by $\Delta {\rm [Fe/H]} \sim 0.5$~dex at the low metallicity end. Although
possible, the low-resolution spectroscopic values are less likely the source of the
problem, given the extensive tests that have been applied in their validation
\citep[see][]{lee:08a,lee:08b,allende:08}. Nevertheless, the relative
metallicity comparison would be robust for stellar populations with the same binary
fraction, even if there is an offset with the SSPP.

In the following analysis, we adjusted our ${\rm [Fe/H]_{phot}}$ to be on the same
scale as the SSPP results by deriving a piecewise linear fit to the median
${\rm [Fe/H]_{phot}}$ as a function of metallicity (solid lines in Fig.~\ref{fig:cluster}).
This adjustment implicitly assumes the same fraction of unresolved binaries applies
to both the SSPP and the field halo samples, if binaries are solely
responsible for the difference between the ${\rm [Fe/H]_{phot}}$ and the SSPP results.
No statistically significant trend was found in the systematic difference between
${\rm [Fe/H]_{phot}}$ and ${\rm [Fe/H]_{spec}}$ for various sets of the binned data with
different magnitude and color ranges.

\section{Results}

\begin{figure}
\epsscale{2.4}
\plottwo{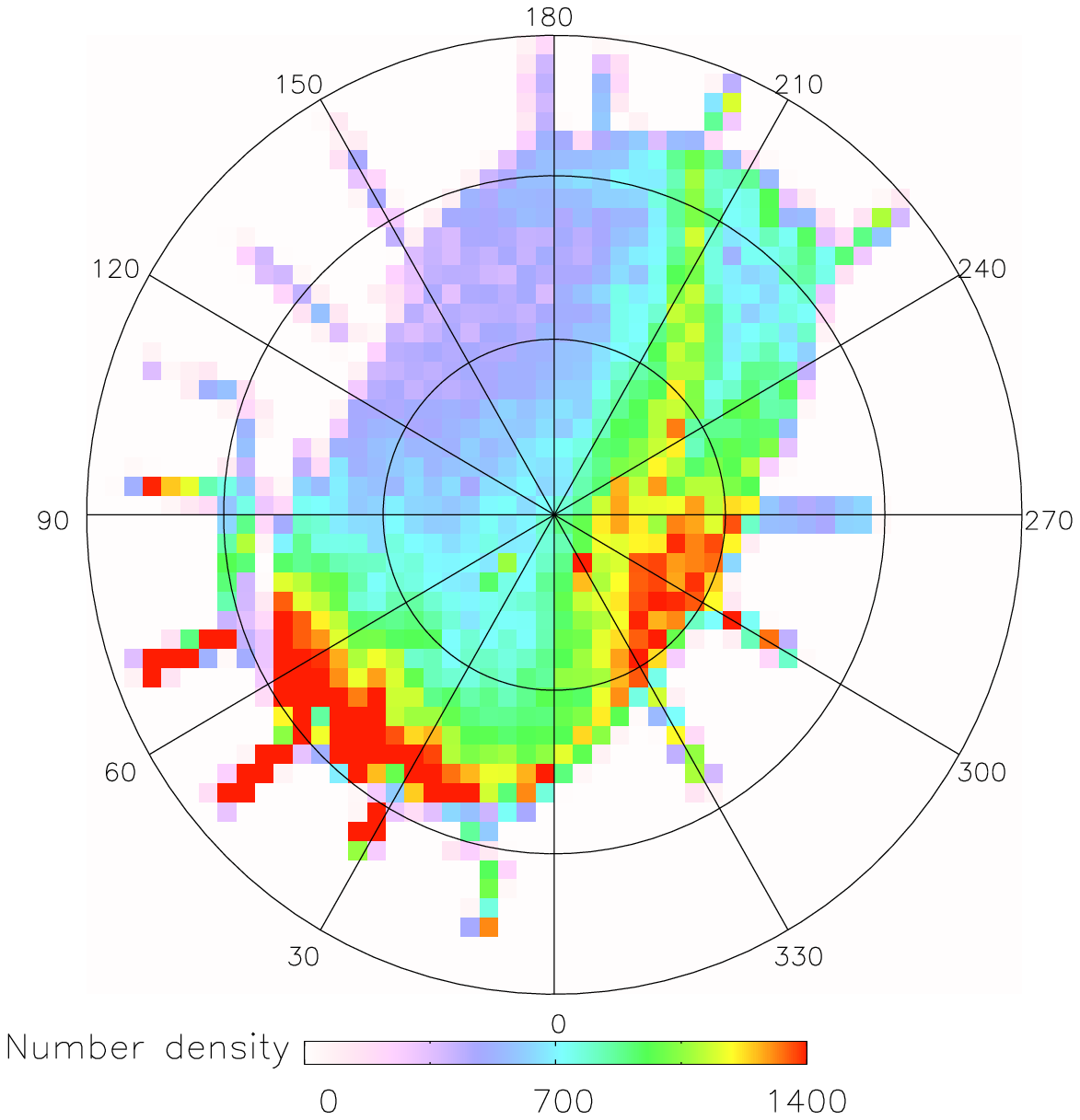}{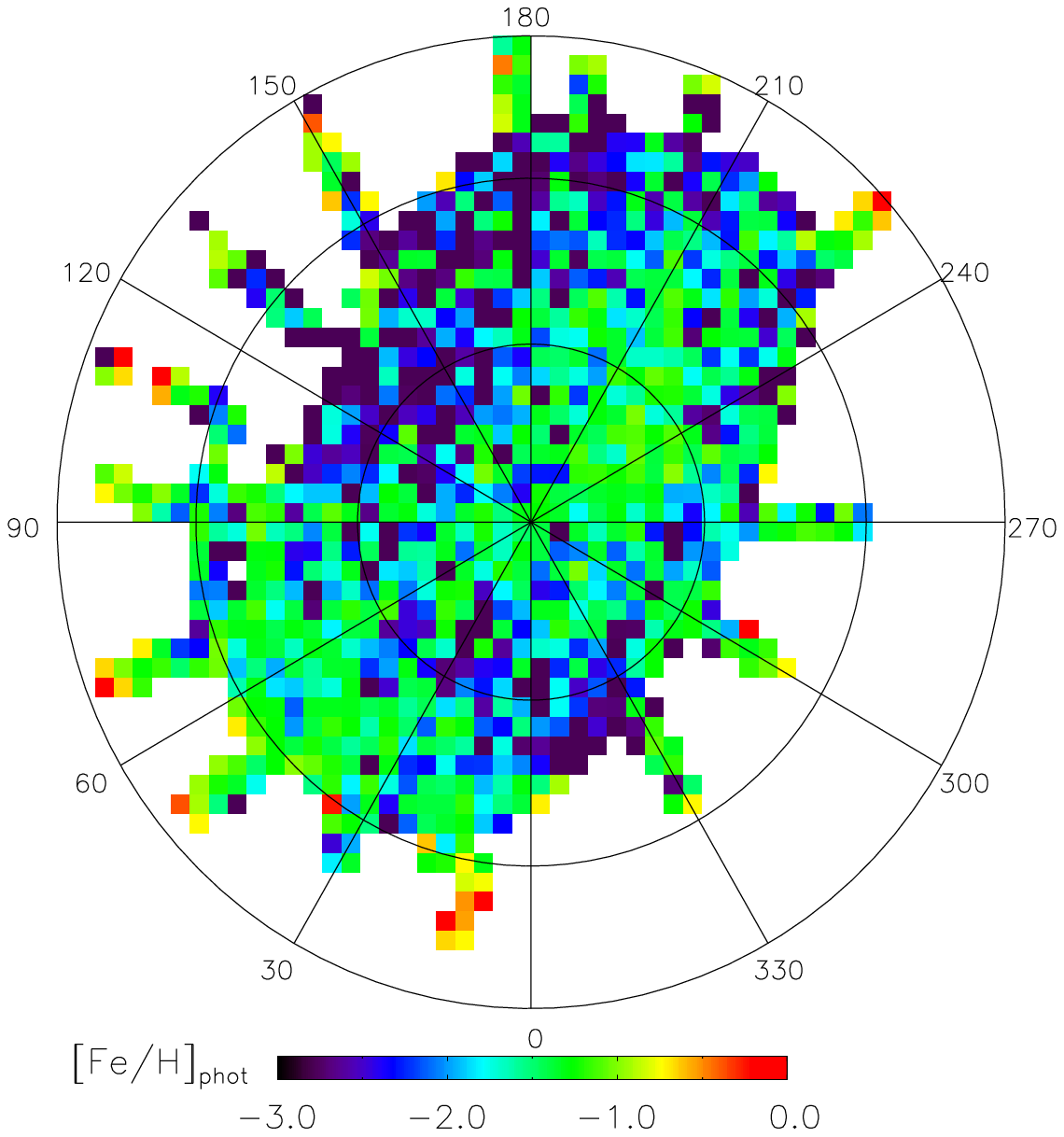}
\caption{{\it Top}: Number density of stars at distances from $\sim10$~kpc to
$\sim20$~kpc from the Sun in the Lambert projection of the Galactic coordinates.
The North Galactic pole is at the center, and the Galactic Center is to the
bottom. Concentric circles represent $b = 0\arcdeg$, $30\arcdeg$, and $60\arcdeg$,
respectively. The VOD is the feature seen at $(l,b) \sim (300\arcdeg,70\arcdeg)$.
{\it Bottom}: Median metallicity of the same stars as in the {\it top} panel. The
median occupancy of each pixel is $544$ for both maps.
\label{fig:halo}}
\end{figure}

Figure~\ref{fig:halo} shows the distribution of stellar number density ({\it top})
and photometric metallicity ({\it bottom}) for 740,658 stars detected in SDSS.
The maps are Lambert equal-area projections of the Northern Galactic hemisphere,
and the distributions are projected as seen from the Sun (i.e., a view from the
inside of the Galaxy). The North Galactic Pole is located at the center, and the
direction of the Galactic Center is toward the bottom in each panel. Each pixel
has an area of $12.96\ \deg^2$. To avoid any possible bias, we restricted our
sample to stars with $0.3 \leq (g - r)_0 \leq 0.4$ and the same color-color cut
as those for the SSPP comparison sample. We used stars with
$15.0 \leq (m - M)_0 \leq 16.5$ ($10 \leq d \leq 20$~kpc).
Although a distance modulus was derived from the $\chi^2$ minimization for each
star, we used a 12.6~Gyr old model with ${\rm [Fe/H]} = -1.6$ for all of the stars
to bracket the distance range in the sample. This was because the uncertainty in
${\rm [Fe/H]_{phot}}$ for individual stars is large enough that the strong
correlation between ${\rm [Fe/H]_{phot}}$ and distance could lead to a biased result.

The feature at $(l,b) \sim (300\arcdeg,70\arcdeg)$ is the VOD \citep{juric:08},
and the long strip that crosses the sky from $(l,b) \sim (210\arcdeg,30\arcdeg)$
to the VOD is the leading tidal tail of Sgr. The feature at $(l,b) \sim (40\arcdeg,
40\arcdeg)$ is the Hercules-Aquila Cloud, another large-area overdensity of halo
stars discovered from SDSS \citep{belokurov:07}.

The median metallicity of stars in the VOD area at $270\arcdeg \leq l \leq 330\arcdeg$
and $60\arcdeg \leq b \leq 70\arcdeg$ is ${\rm [Fe/H]} = -2.0\pm0.1$ from the
metallicity map in Figure~\ref{fig:halo}, where the error is from a pixel-to-pixel
dispersion. The photometric zero points in SDSS vary at the $\sim2\%$ level over
a $\sim10\arcmin$ scale along the scan line \citep{an:08}. They are also known to vary
at the same $\sim2\%$ level over a larger angular scale along the stripe
(a $2.5\arcdeg$ wide SDSS stripe typically runs from the $1^{st}$ to the $3^{rd}$
Galactic quadrant), and from one strip to the other \citep{ivezic:03,ivezic:04}.
Our hope is that these components are properly averaged out over a large area in
the sky, such as the solid angle covered by the VOD.

Field halo stars located at the mirrored position ($30\arcdeg \leq l \leq 90\arcdeg$,
$60\arcdeg \leq b \leq 70\arcdeg$) exhibit ${\rm [Fe/H]} = -1.9\pm0.1$.
Although half of the stars in the direction toward Virgo are likely associated
with a progenitor dwarf galaxy or a tidal stream, the median [Fe/H] value
essentially remains unchanged; it is the same as that for the field halo stars
within the precision of the technique.

It is perhaps of interest that the field-star
metallicities are as low as they appear to be, as \citet{carollo:07} have
argued that the peak metallicity of the outer-halo population is ${\rm [Fe/H]} = -2.2$,
and that this component is expected to dominate over the more metal-rich inner-halo
population (with a peak metallicity at ${\rm [Fe/H]} = -1.6$) at Galactocentric
distances greater than $15$~kpc -- $20$~kpc.
Note that metallicity estimates at the second Galactic quadrant
are even lower than ${\rm [Fe/H]} \approx -2$.

\begin{deluxetable}{lcc}
\tablewidth{0pt}
\tablecaption{Error budget in ${\rm [Fe/H]_{phot}}$ for the VOD\label{tab:sys}}
\tablehead{
  \colhead{Source of Error} &
  \colhead{$\Delta$ Quantity} &
  \colhead{$\Delta {\rm [Fe/H]_{phot}}$}
}
\startdata
Internal                & \nodata      & $\pm0.1$ \nl
Age                     & $\pm20\%$    & $\pm0.3$ \nl
Dust extinction         & $\pm20\%$    & $\pm0.1$ \nl
$[\alpha{\rm /Fe}]$     & $\pm0.1$~dex & $\pm0.3$ \nl
Contamination by giants & \nodata      & $-0.1$   \nl
Alternative approach    & \nodata      & $-0.1$   \nl
Total (systematic)      & \nodata      & $\pm0.5$
\enddata
\tablecomments{These estimates are slightly different from those for the SSPP
sample (\S~2.3) due to different color ranges used.}
\end{deluxetable}

In Table~\ref{tab:sys} we list the sources of several systematic errors and their
contributions to errors in ${\rm [Fe/H]_{phot}}$ for the VOD. The effects of
the age, extinction, and $[\alpha/{\rm Fe}]$ were tested by constructing similar
${\rm [Fe/H]_{phot}}$ maps to Figure~\ref{fig:halo} with different parameters in
the models. Our photometric technique is valid only for main-sequence stars,
but giants constitute approximately $10\%$ of the stars in the sample \citep[see][]{an:08}. Since
photometry alone cannot be used to adequately distinguish giants from dwarfs,
we estimated a bias due to the presence of giants using photometry for a sample of
nearby globular clusters \citep{an:08}. In Table~\ref{tab:sys} we also list the
error from an alternative approach where we
used a median difference in distance modulus in each pixel of Figure~\ref{fig:halo}
to determine ${\rm [Fe/H]_{phot}}$. The total systematic error is the quadrature sum
of all of the error contributions. The relative metallicity comparison is more robust,
if stellar populations in the field halo and the VOD have the same age and [$\alpha$/Fe]
distributions: the difference in ${\rm [Fe/H]_{phot}}$ remains within
$\Delta {\rm [Fe/H]_{phot}} \la 0.2$.

\section{Discussion}

This initial application of the photometric metallicity technique
demonstrated that there is no metallicity difference between the field
halo stars and those in the VOD within the precision of the method. Our
estimate can be compared with previous measurements for
a handful of RR Lyrae variables that are likely associated with the VOD
\citep{duffau:06,vivas:08,prior:09}. These estimates range from ${\rm [Fe/H]} = -1.55$
to $-1.95$, based on the pseudo-equivalent width of the \ion{Ca}{2}~K line.
From principal axes on the $u - g$ vs.\ $g - r$ diagram, \citet{juric:08} argued
that the VOD metallicity is lower than that of thick-disk stars and similar to that
of halo stars.

It is tempting to place the VOD in the observed trend of the luminosity-metallicity
relation among surviving dwarf galaxies in the Local Group \citep[e.g.,][]{grebel:03,
munoz:06,kirby:08b}. If we take our median ${\rm [Fe/H]}$ from Figure~\ref{fig:halo}
at $M_V \sim 10$ \citep{juric:08,prior:09}, the VOD follows this trend, supporting
the idea that [Fe/H] can serve as a luminosity indicator for an accreting dwarf
galaxy, in the process of building up the stellar halos of large spiral galaxies
like the Milky Way \citep[e.g.,][]{johnston:08}. However, this should be taken with
caution, as previous studies often report average abundances for dwarf galaxies
rather than median values.

Future imaging surveys, such as the the Large Synoptic Survey Telescope
\citep[LSST;][]{ivezic:08b} will use similar photometric bandpasses as those in SDSS,
providing even deeper (and {\it far} more accurate) photometric data than SDSS over
a larger fraction of the sky. Our photometric metallicity method will be useful to
exploit these databases for understanding the chemical evolution of progenitor dwarf
galaxies that are identified, as well as for the bulk populations of field stars.

\acknowledgements

We thank James Bullock, \v{Z}eljko Ivezi\'{c}, Heather Morrison, and Katie Schlesinger
for useful discussions.
T.C.B. and Y.S.L. acknowledge partial funding of this work
from grant PHY 08-22648: Physics Frontiers Center/Joint Institute for Nuclear
Astrophysics (JINA), awarded by the U.S. National Science Foundation.
Funding for the SDSS and SDSS-II has been provided by the Alfred P.\ Sloan
Foundation, the Participating Institutions, the National Science Foundation,
the U.S.\ Department of Energy, the National Aeronautics and Space Administration,
the Japanese Monbukagakusho, the Max Planck Society, and the Higher Education
Funding Council for England. The SDSS Web Site is http://www.sdss.org/.

The SDSS is managed by the Astrophysical Research Consortium for the Participating
Institutions. The Participating Institutions are the American Museum of Natural
History, Astrophysical Institute Potsdam, University of Basel, University of Cambridge,
Case Western Reserve University, University of Chicago, Drexel University, Fermilab,
the Institute for Advanced Study, the Japan Participation Group, Johns Hopkins University,
the Joint Institute for Nuclear Astrophysics, the Kavli Institute for Particle
Astrophysics and Cosmology, the Korean Scientist Group, the Chinese Academy of
Sciences (LAMOST), Los Alamos National Laboratory, the Max-Planck-Institute for
Astronomy (MPIA), the Max-Planck-Institute for Astrophysics (MPA), New Mexico
State University, Ohio State University, University of Pittsburgh, University
of Portsmouth, Princeton University, the United States Naval Observatory, and
the University of Washington.

\end{document}